\documentclass{aa}
\usepackage{natbib}
\usepackage{amsmath}
\usepackage[english]{babel}
\usepackage{latexsym}
\usepackage{graphicx}

\newcommand{\plotwd}{8.6cm}

\newcommand{\beal}{\begin{align}}
\newcommand{\bsub}{\begin{subequations}}
\newcommand{\esub}{\end{subequations}}

\newcommand{\Abst}[1]{\,#1}
\newcommand{\id}{{\,\rm d}}
\newcommand{\kB}{k_{\rm B}}
\newcommand{\pot}[2]{#1 \times 10^{#2}}
\newcommand{\xe}{x_{\rm e}}
\newcommand{\lesssim}{\mathrel{\hbox{\rlap{\hbox{\lower4pt\hbox{$\sim$}}}\hbox{$<$}}}}
\newcommand{\gtrsim}{\mathrel{\hbox{\rlap{\hbox{\lower4pt\hbox{$\sim$}}}\hbox{$>$}}}} 
\newcommand{\change}[1]{{#1}}
\begin{document}

\titlerunning{Induced two-photon decay and the rate of hydrogen recombination}

\title{Induced two-photon decay of the 2s level and the \\ rate of cosmological
  hydrogen recombination}

\author{J. Chluba\inst{1} \and R.A. Sunyaev\inst{1,2}}

\institute{Max-Planck-Institut f\"ur Astrophysik, Karl-Schwarzschild-Str. 1,
86740 Garching bei M\"unchen, Germany
\and Space Research Institute, Russian Academy of Sciences, Profsoyuznaya 84/32 Moscow, Russia}

\offprints{J. Chluba, \\ \email{jchluba@mpa-garching.mpg.de}}
\date{Received 5 August 2005/ Accepted 2 September 2005}

\abstract {Induced emission \change{due to the presence of soft CMB photons}
  slightly increases the two-photon decay rate of the 2s level of hydrogen
  defining the rate of cosmological recombination. This correspondingly
  changes the degree of ionization, the visibility function and the resulting
  primordial temperature anisotropies and polarization of the CMB on the
  percent level.  These changes exceed the precision of the widely used {\sc
    Cmbfast} and {\sc Camb} codes by more than one order of magnitude and can
  be easily taken into account.
 }

\maketitle

\section{Introduction}
\label{sec:Intro}
One of the key processes for the formation of the primordial temperature
fluctuations of the cosmic microwave background (CMB) is the recombination of
hydrogen, which at redshifts $z\sim 1100$ makes the Universe transparent for
CMB photons. The time and duration of recombination directly influence the
characteristics of the CMB anisotropies.
Today, line of sight Boltzmann codes like {\sc Cmbfast}
\citep{Seljak1996CMBFAST} and {\sc Camb} \citep{Lewis2000CAMB} are routinely
used to calculate the power spectrum of the primordial temperature
anisotropies with inclusion of many different physical processes in the early
Universe. A precision of the solution to $\sim 0.1\%$ within the assumed
models up to multipoles $l\sim 3000$ is reached.
This level of precision is now becoming necessary with the advent of
space missions like {\sc Wmap}, {\sc Planck} and ground-based experiments such
as {\sc Act} and {\sc Spt}, which will allow us to measure the CMB temperature
and polarization anisotropies with unprecedented accuracy and thereby open a
possibility to determine the key parameters of the Universe with high
precision.

\change{Recently \citet{Dubrovich2005} have included the two-photon decays of
  high levels of neutral hydrogen and helium in their calculation of the
  recombination rates, in the case of hydrogen yielding corrections on the
  level of a few percent and a significant acceleration of helium
  recombination.}
\citet{Leung2004} have included the softening of the matter
equation of state due to the transition from completely ionized to neutral
matter and found that the CMB temperature and polarization power spectra are
affected on the level of some percent at large multipoles. As they pointed out,
these corrections exceed the level of cosmic variance at multipoles $l\gtrsim
1000$ and therefore should be taken into account in high accuracy analysis of
the CMB data.
In this paper we discuss an additional physical process that changes
the ionization degree of hydrogen in the Universe at any given moment of
recombination on the level of a few percent.

It is generally accepted that the rate of recombination is mainly controlled
by the two-photon decay of the metastable 2s level of hydrogen
\citep{ZeldKurtSun1968orig, Peebles1968, Seager1999Apj, Seager2000ApjS}. Here
we discuss the influence of the simulated \change{two-photon emission
  \citep[e.g.  see][ p. 229]{Berestetskii1971}} due to the presence of the low
frequency photons of the CMB blackbody radiation. Below we present a simple
calculation for the change of the two-photon decay rate of hydrogen, $A_{\rm
  2s1s}$, with redshift and shortly discuss the consequences for the
visibility function and CMB temperature and $E$-mode polarization power
spectrum.

\section{Induced $\text{2s}\rightarrow \text{1s}$ two-photon transition of
  hydrogen during recombination}
\label{sec:Rate}
\change{Based on the pioneering work of \citet{Goeppert-Mayer1931}} the rate
for the $\text{2s}\rightarrow \text{1s}$ two-photon transition of hydrogen,
assuming no ambient photon field, has been calculated and discussed many times
using different approaches \citep{Breit1940, Kipper1950, Spitzer1951,
  Goldman1981, Goldman1989}.  \change{Recently} \citet{Labzowsky2005} gave a
value of $A_{\rm 2s1s}=8.2206\,\text{s}^{-1}$ for the total transition rate,
which to 0.1\% agrees with a more simple calculation based on the method used
by \citet{Spitzer1951}.
In this paper we use this simplified approach and include induced effects in
the calculation of the hydrogen two-photon decay rate within the context of
\change{cosmological} hydrogen recombination. A similar calculation can be used to
include the effects of induced two-photon decay for helium recombination.

The total transition rate with no ambient radiation field can be given by
\beal
\label{eq:totA2s1s}
A_{\rm 2s1s}=\frac{A_0}{2}\int^1_0 \phi(y)\id y
\Abst{,}
\end{align}
with $A_0=9\alpha^6 c R/2^{10}\approx 4.3663\,\text{s}^{-1}$, where $\alpha$
is the fine structure constant, $c$ is the speed of light and $R$ is the
Rydberg constant for hydrogen. In equation \eqref{eq:totA2s1s} $\phi(y)\id y$
is proportional to the probability of emitting one photon at frequency
$y=\nu/\nu_0$ in the range $\id y=\id \nu/\nu_0$, where $\nu_0$ is the
frequency of a Lyman-$\alpha$ photon, with energy $\sim 10.2\,$eV, while the
second photon is emitted at $y'=1-y=[\nu_0-\nu]/\nu_0$. The factor of $1/2$ is
required since there are two photons and each pair is counted twice. The
function $\phi(y)$ is defined in the paper of \citet{Spitzer1951} (cf. Eq. 3),
which nowadays can be easily calculated numerically.

\begin{figure}
\centering
\includegraphics[width=\plotwd]
{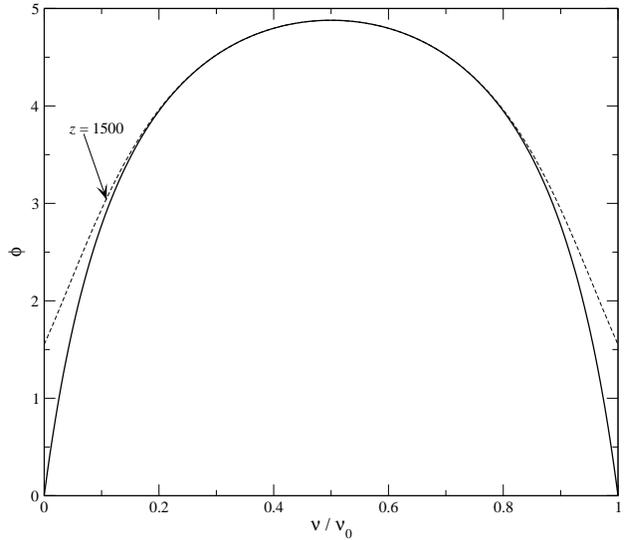}
\caption{Two-photon 2s decay of hydrogen: the solid line shows the two-photon
  probability distribution, $\phi(y)$, as given in \citet{Spitzer1951}
  assuming no ambient radiation field. In contrast to this, the dashed line
  includes the effects of induced emission due to the presence of the CMB at a
  redshift of $z=1500$.}
\label{fig:one}
\end{figure}

In the presence of an ambient radiation field with occupation number $n(\nu)$
the total transition rate is given by the expression
\beal
\label{eq:totA2s1s_ind}
A^{\rm ind}_{\rm 2s1s}=\frac{A_0}{2}\int^1_0 \phi(y)[1+n(\nu)][1+n(\nu_0-\nu)]\id y
\Abst{,}
\end{align}
where the factors of $1+n$ account for the effect of Bose-bunching. In the
context of recombination the radiation field is given by a blackbody spectrum
for which the occupation number is $n(\nu)=1/[e^{h\nu/\kB T}-1]$, with the
photon temperature $T=T_0(1+z)$, where $T_0=2.725\pm 0.001\,$K
\citep{Fixsen2002}.  The relation between $ h\nu/\kB T\equiv x$ and $y$ is
given by
\beal
\label{eq:x2y}
x=\frac{43455}{1+z} \,y\stackrel{\stackrel{z=1100}{\downarrow}}{\approx} 40\,y
\Abst{.}
\end{align}
Figure \ref{fig:one} shows the comparison of the integrands of
\eqref{eq:totA2s1s} and \eqref{eq:totA2s1s_ind}. Due to induced effect the
probability of emitting one soft photon ($y\sim 0$) and at the same time the
other close to the Lyman-$\alpha$ frequency ($y\sim 1$) is enhanced.
This enhancement due to the change of $x$ with time depends on the redshift.
At sufficiently small redshifts, induced effects become negligible.

One can examine the behavior of $\phi(y)$ for $y\ll 1$ more closely by using
the analytic fit as given by \citet{Nussbaumer1984}. With this, one obtains
\beal
\label{eq:phi_appr}
\phi(y)=C\left[w(1-4^\gamma\,w^\gamma)+\alpha\,w^{\beta+\gamma}\,4^\gamma\right]
\Abst{,}
\end{align}
where $w=y[1-y]$, $C=46.26$, $\alpha=0.88$, $\beta=1.53$ and $\gamma=0.8$
(note that \change{in \citet{Nussbaumer1984}} the normalization constant $C$ was
defined differently). With this the integrands of equations
\eqref{eq:totA2s1s} and \eqref{eq:totA2s1s_ind} around $y\sim 0$ can be
expressed as
\bsub
\label{eq:phi_approx}
\beal
\phi(y)&\approx C\,y\left[1-3.03\,y^{0.8}\right]
\\
\phi^{\rm ind}(y)&\approx \frac{C}{\kappa}\left[1-3.03\,y^{0.8}+\frac{\kappa-2}{2}\,y \right]
\Abst{,}
\end{align}
\esub
respectively, where here we introduced 
\beal
\kappa=\frac{43455}{1+z}
\Abst{.}
\end{align}
\change{These approximations are accurate within a few percent for $y\leq
  0.05$ in the redshift range $z=1000-1500$.}

Here it is important that $\phi(y)$ vanishes as $\phi(y)\sim y$ for
$y\rightarrow 0$. Since in this limit $n(\nu)\sim 1/y$, the product
$\phi(y)n(\nu)$ is finite. Therefore it is not necessary to introduce a low
frequency cut-off in the integral~\eqref{eq:totA2s1s_ind}.

From \eqref{eq:phi_approx} one can also deduce the contribution of the low
frequency part to the total two-photon decay rate. Integrating from zero up to
$y_{\rm m}$ one finds
\bsub
\label{eq:A_approx}
\beal
\Delta A_{\rm 2s1s}
&\approx 50.5\,y_{\rm m}^2\left[1-2.16\,y_{\rm m}^{0.8}\right]\,{\rm s^{-1}}
\\
\Delta A_{\rm 2s1s}^{\rm ind}
&\approx \frac{101}{\kappa}\,y_{\rm m}\left[1-1.68\,y_{\rm m}^{0.8}
  +\frac{\kappa-2}{4}\, y_{\rm m} \right]\,{\rm s^{-1}}
\Abst{.}
\end{align}
\esub
\change{Here $\Delta$ indicates that only part of the integrals
  \eqref{eq:totA2s1s} and \eqref{eq:totA2s1s_ind}, i.e. for $y\in [0,y_{\rm
    m}]$ have been taken.}  Due to the symmetry of $\phi(y)$ around $y=1/2$
one can apply the approximations \eqref{eq:phi_approx} and \eqref{eq:A_approx}
also to the case $y\rightarrow~1$ by simply replacing $y\rightarrow 1-y$.

If one assumes that photons very close to the center of the Lyman-$\alpha$
line cannot escape then this reduces the contribution of the two-photon decay
to the effective rate of recombination, since the trapped photons prevent the
corresponding recombination.
\change{As an example, if at redshift $z=1100$ photons within 1\% (5\%) of the central
frequency of the Lyman-$\alpha$ line cannot escape then the effective
two-photon decay rate (i.e. the rate used in the calculation of the
recombination history) is smaller by $\sim 0.06\%$ ($\sim 1.23\%$) for the
standard calculation of $A_{\rm 2s1s}$ and by $\sim 0.33\%$ ($\sim 2.01\%$) for
the calculation including both spontaneous and induced effects.

As will be shown below, in general the contributions due to induced effects
are at the level of some percent themselves. Hence, this estimate shows that
the corrections to the two-photon decay rate we are discussing here can only
be considered accurate if the more energetic photons from the 2s two-photon
decay lying within less than $\sim 0.1-1\%$ of the Lyman-$\alpha$ line center
are trapped.
At redshift $z\sim 1100$ the Doppler width of the Lyman-$\alpha$ line is
$\Delta \nu_{\rm D}\sim \pot{2.3}{-5}\,\nu_0$. Our computations of the
Lyman-$\alpha$ photon escape in the distant wings show that there is no
significant diffusion back to the line center beyond a few ten to hundred
Doppler widths. Therefore the aforementioned condition should be easily
fulfilled, since at $z=1100$ a 1\% distance from the line center corresponds
to $\sim 435$ Doppler widths. However, even if every photon within 1\% of
the line center is unable to escape, this would only affect the
results obtained below by about ten percent.}
One should note that the expected corrections due to the Lamb-shift are
much smaller.

\begin{figure}
\centering
\includegraphics[width=\plotwd]
{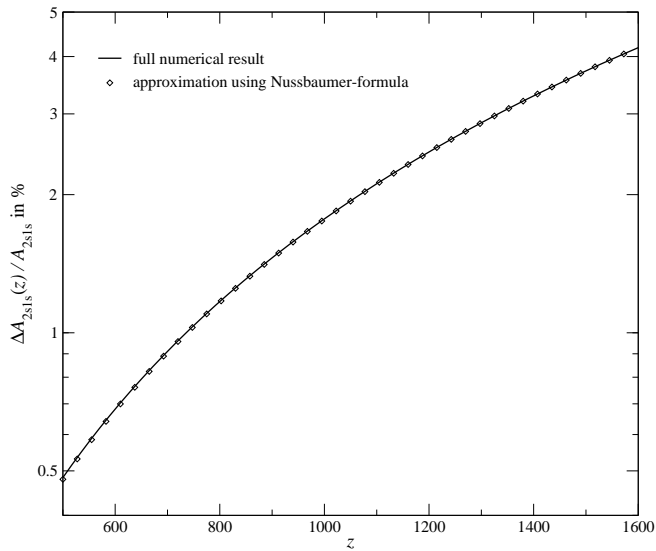}
\caption
{Redshift dependence of the relative change of the total two-photon transition
  rate of hydrogen, $[A^{\rm ind}_{\rm 2s1s}-A_{\rm 2s1s}]/A_{\rm 2s1s}$ with
  $A_{\rm 2s1s}=8.2206\,\text{s}^{-1}$, in the context of recombination using
  Eq. \eqref{eq:totA2s1s_ind}. \change{In addition the results obtained with
    the simple analytic approximation \eqref{eq:phi_appr} for $\phi(y)$ are
    shown.}}
\label{fig:two}
\end{figure}
In Figure \ref{fig:two} we present the redshift dependence of the relative
change of the total two-photon transition rate of hydrogen in the context of
recombination using equation \eqref{eq:totA2s1s_ind}. A simple fit to this
function, which within $\lesssim 1$\% accuracy is applicable in the redshift
range $500\leq z \leq 2500$, can be given by the expression
\beal
\label{eq:DA_appr}
\frac{A^{\rm ind}_{\rm 2s1s}-A_{\rm 2s1s}}{A_{\rm 2s1s}}
&=\pot{1.181}{-3}\,\chi
+\pot{2.177}{-2}\,\chi^2
\nonumber \\
&\qquad
-\pot{1.958}{-3}\,\chi^3
\Abst{,}
\end{align}
with $\chi=(1+z)/1100$ \change{and where $A_{\rm 2s1s}$ and $A^{\rm ind}_{\rm
    2s1s}$ are given by equations \eqref{eq:totA2s1s} and
  \eqref{eq:totA2s1s_ind}, respectively.} The main correction scales as
$\propto (1+z)^2$.  The total correction exceeds the percent level for
$z\gtrsim 700$.  
\change{In addition the results obtained with the simple analytic
  approximation \eqref{eq:phi_appr} for $\phi(y)$ are shown. They agree very
  well with the full numerical result but were obtained with much less
  numerical effort.}

The rate for the inverse process can be found from equation
\eqref{eq:totA2s1s_ind} using the principle of detailed balance. In
thermodynamic equilibrium the ratio of the population of the 1s and 2s levels
is given by $N_{\rm 2s}/N_{\rm 1s}=\exp(-E_{21}/\kB T)$, where
$E_{21}=10.2\,$eV. Therefore the rate for the inverse process is given by
$A_{\rm 1s2s}=A^{\rm ind}_{\rm 2s1s}\exp(-E_{21}/\kB T)$. However, during the
period of recombination, which is most relevant for the CMB anisotropies, the
inverse process is negligibly small.

\section{Changes of the ionization fraction, visibility function and power spectra}
\label{sec:mod_CMBFAST}
It is straightforward to include the redshift dependence of the 2s two-photon
transition rate into the {\sc Cmbfast} code using the approximation
\eqref{eq:DA_appr}.  One only has to replace the standard two-photon decay
rate $A_{\rm 2s1s}$ by $A_{\rm 2s1s}^{\rm ind}(z)$.
With this one can calculate the changes in the ionization fraction, the
visibility function, $\mathcal{V}(z)=\exp(-\tau)\id\tau/\id \eta$
\citep{SunyZeld1970} and the temperature and $E$-mode polarization power
spectra for the {\sc Wmap} concordance model \citep{WMAP_params}. \change{Above, $\tau$
is the Thomson optical depth and $\eta$ is the conformal time.}

\begin{figure}
\centering
\includegraphics[width=\plotwd]
{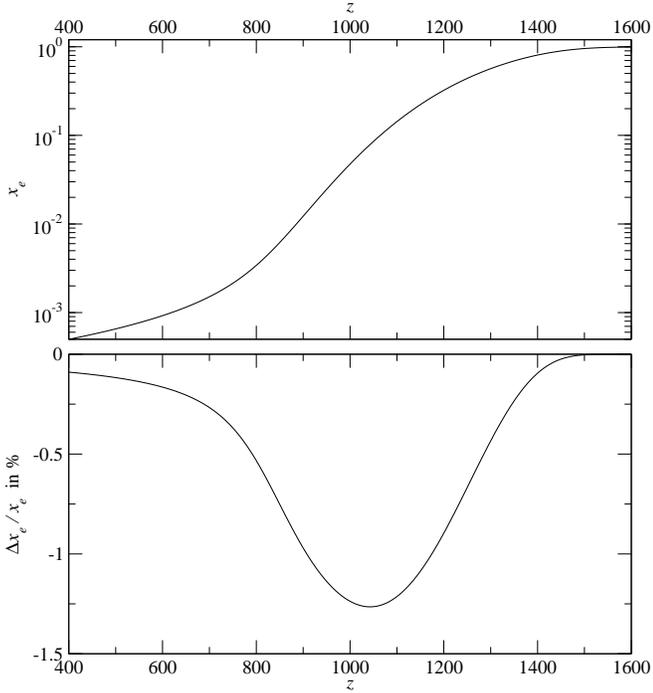}
\caption
{Ionization fraction, $\xe=N_{\rm e}/N_{\rm H}$, and the relative change due
  to the inclusion of induced two-photon emission for the {\sc Wmap}
  concordance model. Here $N_{\rm e}$ and $N_{\rm H}$ are the free electron
  and hydrogen number densities, respectively.}
\label{fig:three}
\end{figure}
In Figure \ref{fig:three} we have presented the redshift dependence of the
ionization fraction and the corresponding change due to the inclusion of
induced two-photon emission for the {\sc Wmap} concordance model using a
modified version of {\sc Cmbfast}. The ionization fraction is affected by a
few percent with a maximal relative difference of $\sim -1.3\%$ at $z\sim
1050$. Because the total two-photon decay rate is slightly higher than in the
standard calculation, recombination occurs a bit faster.

\begin{figure}
\centering
\includegraphics[width=\plotwd]
{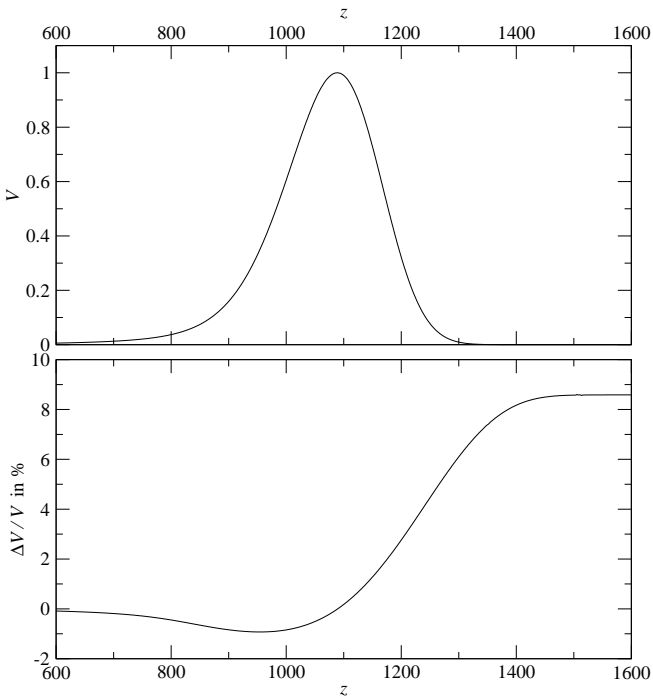}
\caption
{Visibility function, $\mathcal{V}(z)=\exp(-\tau)\id\tau/\id \eta$, and the
  relative change due to the inclusion of induced two-photon emission for the
  {\sc Wmap} concordance model. The amplitude of $\mathcal{V}(z)$ is normalized,
  such that the maximum is 1.}
\label{fig:four}
\end{figure}

In Figure \ref{fig:four} we show the results obtained for the visibility
function. At redshifts much below the maximum at $z_{\rm dec}=1089\pm 1$
\citep{WMAP_params} the visibility function is affected by less than one
percent, whereas for $z\gg z_{\rm dec}$ the change reaches $8\%$. The position
of the maximum and the width of the visibility function are both affected on a
level below one percent.

\begin{figure}
\centering
\includegraphics[width=\plotwd]
{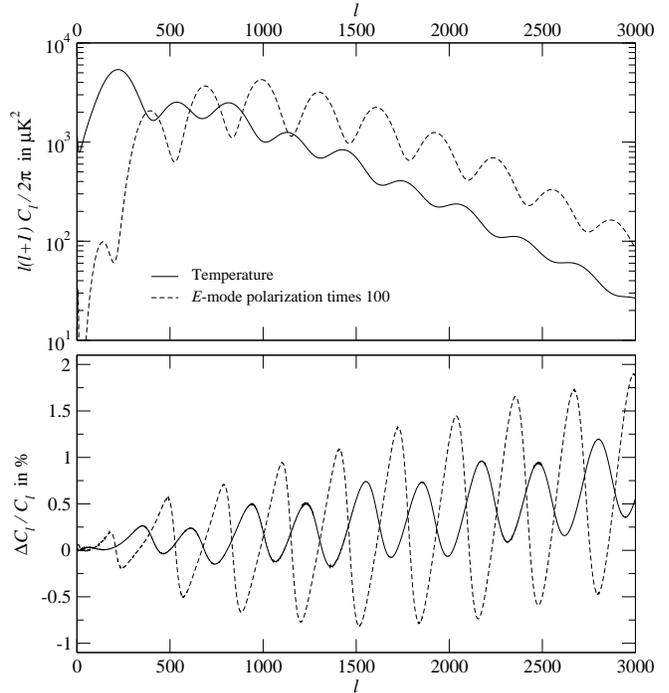}
\caption
{Temperature and $E$-mode polarization power spectra and their relative change
  due to the inclusion of induced two-photon emission for the {\sc Wmap}
  concordance model.}
\label{fig:five}
\end{figure}

In Figure \ref{fig:five} we give the results obtained for the temperature and
$E$-mode polarization power spectra. Again the changes due to induced
two-photon emission are on the level of a few percent, with an increase towards
smaller scales. The amplitude of the change of the polarization power spectrum
is roughly twice that of the temperature power spectrum.

\section{Conclusion}
Due to induced two-photon decay of the hydrogen 2s level, the rate of
recombination is increased on the level of a few percent around the maximum of
the visibility function. This increase results in changes of the ionization
fraction, the visibility function and the temperature and polarization power
spectra by a few percent. These changes can be easily taken into account for
future high accuracy analysis of CMB data using the approximation
\eqref{eq:DA_appr}.
Induced two-photon decay would similarly influence the recombination of HeII
and HeIII, but in that case the effects on the CMB power spectra are expected
to be extremely small.

\acknowledgements{The authors wish to thank the referee, Prof. Dubrovich, for
  his comments on the manuscript. We acknowledge the use of {\sc
  Cmbfast} for the calculation of the power spectra.}

\bibliographystyle{aa}
\bibliography{Lit}

\end{document}